\documentclass[aps, twocolumn, showpacs]{revtex4-1}
\usepackage{amsmath}
\begin{document}
\title{Physical reality of electromagnetic potentials and the classical limit of the Aharonov-Bohm effect}
\author{S. C. Tiwari \\
Department of Physics, Institute of Science, Banaras Hindu University, Varanasi 221005, and \\ Institute of Natural Philosophy \\
Varanasi India\\}
\begin{abstract}
Recent literature on the Aharonov-Bohm effect has raised fundamental questions on the classical correspondence of this effect and the physical reality of the electromagnetic potentials in quantum mechanics. Reappraisal on Feynman's approach to the classical limit of AB effect is presented. The critique throws light on the significance of quantum interference and quantum phase shifts in any such classical correspondence. Detailed analysis shows that Feynman arguments are untenable on physical grounds and the claim made in the original AB paper that this effect had no classical analog seems valid. The importance of nonintegrable phase factor distinct from the AB phase factor, here termed as Fock-London-Weyl phase factor for the historical reasons, is underlined in connection with the classical aspects/limits. A topological approach incorporating the physical significance of the interaction field momentum is proposed. A new idea emerges from this approach that attributes the origin of the AB effect to the exchange of modular angular momentum. 
 
\end{abstract}
\maketitle

\section{\bf Introduction}

Recently conceptual issues of fundamental physical significance have been raised on the Aharonov-Bohm (AB) effect \cite{1,2}. Nonlocality and/or independent physical reality of the electromagnetic potentials seemingly implied by the AB effect have been questioned and debated in the recent literature \cite{3,4,5,6,7}. The original paper on the AB effect \cite{2} is written with remarkable clarity; to get proper perspective on the current controversy it would be appropriate to emphasize salient features contained in this article following recent historical account \cite{8} and theoretical study on gauge invariance \cite{9}.

Arbitrariness in choosing the potentials in classical electrodynamics embodied in the gauge transformation
\begin{equation}
A^\mu ~ \rightarrow ~ A^\mu - \partial^\mu \chi
\end{equation}
and the consequent invariance of the Lorentz-Maxwell theory unambiguously demonstrate the fact that the potentials are just auxiliary mathematical tools in the classical theory. The principle of gauge invariance in quantum mechanics acquires new significance first recognized by Fock in 1926 \cite{8} since the Schroedinger wave function of the charged particle, let us assume it to be electron, gets multiplied by a local phase factor
\begin{equation}
\psi ~ \rightarrow ~ \psi~exp{({\frac{ie}{\hbar c}}\chi)}
\end{equation}
Though the potentials are needed in canonical formalism of the classical theory they do not appear in the equation of motion; in contrast, the electromagnetic potentials cannot be eliminated in quantum theory of electron interacting with the electromagnetic fields. However gauge invariance and the unobservability of phase factors would seem to deny the physical reality to the potentials even in quantum theory. The ingenuity in the Aharonov-Bohm argument lies in the consideration of loop integrals in the phase of the wavefunction. For example, the most celebrated one given schematically in Figure 2 of \cite{2} for the vector potential defines the AB phase shift to be
\begin{equation}
\Delta \delta = -\frac{e}{c\hbar} \oint {\bf A}.d{\bf x}=-\frac{e}{c\hbar} \Phi
\end{equation}
where the total magnetic flux confined to a small region inaccessible to the interfering electron beams is
\begin{equation}
\oint {\bf A}.d{\bf x}=\int_S ({\bf \nabla}\times{\bf A}).d{\bf S} = \Phi
\end{equation}
This phase shift would manifest as a shift in the whole interference pattern of the electron beams relative to that in the absence of the flux or to that obtained by varying the magnetic flux.

Physical interpretation of the AB phase shift involves two important observations made in \cite{2}. The role of the pure gauge field in the field-free region amounts to a multiply connected space
\begin{equation}
{\bf B} ={\bf \nabla}\times {\bf A} =0  ~ \Rightarrow {\bf A} ={\bf \nabla}\phi
\end{equation}
where $\phi$ is a multi-valued scalar field. Now the electron wavefunction (2) is no longer a single-valued function demanded in quantum mechanics; a novel suggestion is made by the authors to split the electron beam into two components encircling the flux region on opposite sides with the corresponding wavefunctions being single-valued. The phase shift in each beam is calculated in terms of a path-dependent phase factor $\frac{e}{c} \int {\bf A}.d{\bf x}$. If the beams are recombined for quantum interference then the relative phase equals the AB phase shift for a closed path i. e. the expression (3). The ideal AB scheme consists of double-slit interferometer, perfectly shielded magnetic flux confined in a small region behind the wall between two slits making it inaccessible to electrons on both sides, and static vector potential. The main conclusions drawn by the authors could be summarized as follows.

{\bf C1.} Shift in the whole interference pattern due to flux is an observable effect. The AB effect has no classical analog since the quantum wave mechanical nature of electrons is crucial for the interference phenomenon.

{\bf C2.} Topological nature is implied by multiply connected space.

{\bf C3.} Gauge invariance is not violated.

{\bf C4.} The absence of magnetic field on the path of electron beams implies either one postulates nonlocal interaction that conflicts with relativity principle or attributes physical reality to the potentials in the quantum domain. Authors prefer second option \cite{2} stating that,'we are led to regard $A_\mu(x)$ as a physical variable'.

The most important physics issue is whether the ideal AB phenomenon could be realized in the laboratory experiments. Numerous experiments performed over past more than five decades have very nearly implemented the AB scheme, and demonstrated the AB effect. Nevertheless there remains the scope for a genuine doubt regarding the perfect shielding of the flux region \cite{4}. In view of the lack of the quantitative estimates on the empirical data seriously challenging the observed AB effect \cite{10} and the beauty of topological interpretation, the alternatives advocating local interaction of fields receive very little attention. Speculative arguments, however abound in the literature; admittedly philosophical beliefs and thought experiments do have their own importance as exemplified in the current controversy \cite{3,5,6,7}. In such cases the pitfalls involving misinterpretations have to be carefully addressed. A recent experiment using time of flight measurement of electrons in the AB setup \cite{11} seems to rule out classical and semiclassical explanations of AB effect. In contrast, single slit diffraction experiment using ballistic electrons in  quantum point contacts \cite{12} claims the equivalence of quantum formalism a la Aharonov-Bohm phase to the classical picture of electrons under the Lorentz force.  We have pointed out that the claim could be misleading \cite{13}. The classical correspondence discussed in \cite{12} is based on the so called Feynman's thought experiment \cite{14}.

In the light of the categorical assertion C1 made in \cite{2} and the recent experimental results reported in \cite{11,12} it becomes crucial to examine the role of classical limit and classical interpretations of the AB effect. Another equally important issue that has emerged concerns the meaning of nonlocality in relation to C4. Nonlocal exchange of modular momentum as a physical mechanism for the AB effect proposed in 1969 \cite{15} has been recently emphasized and two distinct aspects, namely the continuous and instantaneous ones are proposed in \cite{7,16}. Note that there is no experimental evidence for such a nonlocal process. The aim of the present paper is two-fold: to offer a thorough reappraisal on the classical perspectives and nonlocal issues, and to propose a new physical  mechanism for the AB effect in terms of modular angular momentum exchange. It is shown that Feynman approach proving classical-quantum eqivalence of the AB effect is not just puzzling \cite{13} it is incorrect. The importance of a path dependent phase factor that we term as Fock-London-Weyl (FLW) phase distinct from the topological AB phase is discussed in this context. Recent insights gained on the double slit quantum interference \cite{17,18,19} lead us to understand the controversy on classical limit of the AB phase measurement. The question of AB phase evolution raised in \cite{7,16} is analyzed pointing out subtle difference between geometry and topology of the phenomenon. The effect of pure gauge field is discussed in analogy to angular momentum holonomy suggested for the geometric phases in optics \cite{20}.

The paper is organized as follows. The basic concepts on double slit quantum interference of electrons based on the actual experiments \cite{17,18,19}, and on the role of gauge invariance in quantum mechanics following \cite{8,9} constitute next section. A detailed analysis of Feynman approach to the classical limit of AB effect is presented in Section III. Section IV deals with the classical perspective on the AB phenomenon. Brief review on the nonlocal modular momentum is followed by a suggested new approach to explain AB effect in Section V. Concluding remarks constitute the last section.

\section{\bf Basic Concepts}

Proposed experimental test of the AB phase shift (3) discussed in \cite{2} mentions the use of a solenoid or magnetized whiskers to create confined magnetic flux in the double slit experiment for the coherent electron beams. It is also pointed out that it would be convenient to observe the effect by varying the magnetic flux and that the induced electric field could be made negligible. Authors explicitly state that vector potential is time-independent, therefore, at least in this kind of AB phenomenon their argument invoking relativity in the last section as regards to the objection to local interaction is unjustified. We re-emphasize that the AB effect discussed in \cite{2} is a nonrelativistic quantum phenomenon. To avoid confusions it is necessary to understand the delicate issues of double slit experiment in quantum mechanics and the salient features of gauge invariance.

{\bf A. Double slit quantum interference}

Young's double slit interference experiment in optics, initially considered as a thought experiment for de Broglie matter waves, has been realized in the laboratory experiments for decades. Tonomura et al \cite{17} reported an important experiment demonstrating wave-particle duality and quantum mechanical nature of the observed interference on the electron beams.

Let us briefly recall the elementary considerations of two beam interference based on the division of wave-front of a single beam of light in double slit experiment. The interference pattern consists of equidistant bright and dark bands on the plane of the screen. For small separation d between the slits $S_1$ and $S_2$ and the distance between the slits and the screen L the optical path difference from $S_1$ and $S_2$ to a point P(x,y) on the screen is calculated to be
\begin{equation}
\Delta s=\frac{xd}{L}
\end{equation} 
in the approximation of short wavelength $\lambda_0$ and $d<<L$. The corresponding phase difference is
\begin{equation}
\delta =\frac{2\pi}{\lambda_0} ~ \frac{xd}{L}
\end{equation}
The light intensity maxima and minima for the superposition of the beams occur at $\delta =2m\pi$, $|m| =0,1,2..$ and 
$|m|=1/2, 3/2 ...$ respectively. Fresnel biprism is another method in which refraction divides a beam of light into two coherent components and their superposition resulting into the interference phenomenon.

It is known that one can treat electron optics in analogy to light, and define refractive index for an electron beam passing through a purely electrostatic field; Tonomura et al \cite{17} make use of this in the biprism experiment. Assuming incident electron beam to be a plane wave $e^{ikz}$ propagating in the z-direction, after traversing a region with electrostatic potential V(x,z) it is transformed to
\begin{equation}
\Psi (x,z) = exp ~i[k_z z -\frac{em}{\hbar^2 k_z} \int_{-\infty} ^z V(x,z^\prime)~dz^\prime]
\end{equation}
The details of the numerical values and the approximate potential function can be found in \cite{17}, here we give a short account on their significant results. The actual interference is built from a succession of single electrons over a period of time. Note that it is a single electron wave passing through both slits that forms quantum probability interference. On the screen, a detector records an electron as a localized particle i. e. electron wavefunction collapses to a definite position. It is crucial that which-path information for electron trajectory in the classical sense does not exist; thus quantum mechanical wave nature of an electron is essential for observing interference. Position-sensitive electron counter on the 2-dimensional screen records particle nature of the electron.

Interference experiments in classical optics do not have the enigmatic role of wave-particle duality, of course, single photon quantum optics double slit experiments have similar interpretational issues as noted above for electrons. Einstein-Podolsky-Rosen (EPR) incompleteness argument on the foundations of quantum mechanics in 1935 \cite{21} are no longer philosophical; a large volume of experimental work with the advances in technology throws light on them. The Bohr-Einstein debate could be addressed avoiding mysterious or counter-intuitive descriptions of the past \cite{22}. Two representative experiments \cite{18,19} are discussed here which have bearing on the double slit interference. Historically, Bohr in a detailed response to EPR argument \cite{23} conceived single slit diffraction and double slit interference thought experiments to elucidate his complementarity principle. In a double slit experiment the knowledge of the path of a particle passing through one of the slits would wipe out the interference fringes since the position of the particle is ascertained from a measurement of the momentum transfer to the diaphragm and the Heisenberg uncertainty principle comes into play. Note that in Tonomura et al experiment \cite{17} which-path knowledge is not known since it is only on the screen that particle position is determined. This experiment does not test Bohr's prediction; it just conforms to quantum mechanics.

In Eichmann et al experiment \cite{18} Bohr's assertion is proved though position-momentum uncertainty is not invoked. Two $^{198}Hg^+$ ions trapped in a linear Paul trap act as slits for photons. Internal electronic levels of the ions provide which-path information in a polarization-sensitive detection. A linearly polarized photon scattered from the ions is either $\pi$-polarized or $\sigma$-polarized. In the case of $\pi$-polarized scattered photon the ions' electronic levels remain unchanged and hence the knowledge as to which ion scattered the photon is unknown: one should observe interference pattern that is confirmed in the experiment. If the scattered photon is $\sigma$-polarized, one of the ions undergoes electronic level transition resulting into which-path information: observed disappearance of the fringes in this case validates Bohr's prediction. Authors are careful to mention that correlation between the object and the measuring instrument explains the destruction of the interference fringes since Heisenberg uncertainty relation for position and momentum is not needed.

Schmidt et al experiment \cite{19} attempts to explore momentum transfer to the slits, and to study its effect on the interference. Free floating diatomic ions $HD^+$ act as slits for helium atoms. Both quantum mechanical and semi-classical calculations are carried out by the authors and compared with the experimental observations: only former agrees with the experiments. An interesting study is also reported: the semiclassical model is modified such that the momentum transfer is equally divided between both nuclei in each collision. This calculation gives good agreement with the observations. Authors term this modification as classical analog of coherent momentum transfer. Curiously though the force acts only on one of the scattering centers, the momentum is transferred to both. Its similarity to the process in AB effect deserves attention.

To summarize: classical correspondence of the double slit quantum interference is an intricate issue due to wave-particle duality. 

{\bf B. Gauge invariance}

Gauge invariance in electromagnetism is a standard textbook subject, however there exist subtle points that sometimes get overlooked or unrecognized as noted by Wu and Yang \cite{9}. Motivated by the AB effect the authors review the role of the phase defined by a loop integral
\begin{equation}
\frac{e}{\hbar c} \oint A_\mu dx^\mu
\end{equation}
and the phase factor
\begin{equation}
\Delta_{AB} = exp ~(\frac{ie}{\hbar c} \oint A_\mu dx^\mu)
\end{equation}
emphasizing the fact that though the phase (9) contains more information than the phase factor $\Delta_{AB}$ the additional information is not measurable. They further argue that one is naturally led to the basis for the description of electromagnetism to a path-dependent (or nonintegrable) phase factor
\begin{equation}
\Delta_{P_1P_2} = exp ~ (\frac{i e}{\hbar c} \int_{P_1}^{P_2} A_\mu dx^\mu)
\end{equation}
This quantity as compared to $\Delta_{AB}$ is more generally applicable to the dynamics of a charged particle in an electromagnetic field. Since this aspect is intrinsic to the ideas of Fock, London, and in a more concrete form that of Weyl \cite{8} we suggest that the path-dependent phase factor be termed as FLW phase. This terminology would have the advantage that unnecessary confusion created in the literature by using the term AB phase for expression (11) would be avoided. Moreover it becomes clear that FLW phase factor describes charged particle interaction with the electromagnetic field in a transparent manner. Note that the gauge transformation assumes the form
\begin{equation}
\Delta_{P_1P_2} ~ \rightarrow e^{\frac{ie}{\hbar c} \chi(P_2)}
\Delta_{P_1P_2}e^{-\frac{ie}{\hbar c} \chi(P_1)} 
\end{equation}

The preceding discussion shows that FLW phase factor can be used to obtain Schroedinger wavefunction for an electron in the presence of the electromagnetic field in terms of the free electron wavefunction $\Psi_0$
\begin{equation}
\Psi = exp(-\frac{ie}{\hbar c} \int^x A^\mu dx_\mu) \Psi_0
\end{equation}
Alternatively, the Schroedinger equation for an interacting electron can be derived from the free-particle Schroedinger equation expressing $\Psi_0$ in terms of $\Psi$ from Eq.(13). For example, the free-electron wavefunction in the expression (8) expressed in terms of $\Psi(x,z)$ gives the interaction term $eV$ in the Schroedinger equation: the operator $i\hbar \frac{\partial}{\partial t}$ on $exp i[\frac{em}{\hbar^2 k_z} \int_{-\infty} ^z V(x,z^\prime)~dz^\prime]$ yields $eV$ noting that the integration variable $z^\prime$ can be changed to $dz^\prime=v_z dt^\prime$, where $v_z =\hbar k_z/m$, and the partial time-derivative on the integral finally just gives $V(x,z)$. One can obtain the Schroedinger equation for an electron in arbitrary electromagnetic field from Eq.(13) using the free-particle wave equation
\begin{equation}
(-\frac{\hbar^2}{2m} \nabla^2 +\frac{ie\hbar}{mc} ({\bf A}.{\bf \nabla} +\frac{1}{2} {\bf \nabla}.{\bf A}) +\frac{e^2}{2mc^2} {\bf A}^2 +e \phi)\Psi =i\hbar \frac{\partial \Psi}{\partial t}
\end{equation}
Note that the classical limit makes sense for Eq.(14) as we discuss below.

\section{\bf Feynman's analysis of Aharonov-Bohm effect}

In one of the earliest textbook treatments Feynman presents an expository discussion on the AB effect \cite{14}, and proceeds to raise the question of the classical correspondence of the quantum significance of the vector potential stating that,'...if we look at things on a large enough scale it will look as though the particles are acted on by a force equal to $q{\bf v} \times$ the curl of ${\bf A}$'. The recent experiments \cite{11,12} and the importance of the correspondence principle in the recent discussions on the interpretation of the AB effect \cite{3,6,7} invite attention to the Feynman's analysis for gaining deeper insights. In fact, the conceptual issue raised by Feynman is whether the vector potential is a real physical field. A discussion on the arbitrariness of the potentials in the classical electrodynamics in Section 15-4 leads him to explore the role of ${\bf A}$ in quantum mechanics in Section 15-5. For the sake of clarity and to bring out the importance of Feynman's complete analysis we divide it into three parts.

{\bf Part-A}

Recalling the quantum mechanical wave nature of electrons in the double slit electron interference thought experiment Feynman argues that the effect of electric and magnetic fields on the electron wave manifests in the form of phase changes in the wavefunction. Considering the double slit experiment in the presence of the magnetic field $B$ the phase difference is obtained in terms of the flux of $B$ enclosed by two paths making use of the Stokes theorem
\begin{equation}
\delta = \delta (B=0) - \frac{e}{\hbar c}~ Flux
\end{equation}
Though the vector potential determines the phase change along a trajectory given by
\begin{equation}
Magnetic ~ change ~ in ~ phase =-\frac{e}{\hbar c} \int^x {\bf A}.d{\bf x}
\end{equation}
Feynman states that since it is the magnetic field that appears in the expression (15) it would seem that ${\bf A}$ is an 'artificial construction'.

Remark one : In this part of Feynman's analysis there is no role of AB phase (3) where the line integral of ${\bf A}$ is evaluated in the field-free region. It is easily recognized that Feynman's magnetic change in phase (16) is exactly the path-dependent FLW phase appearing in $\Delta_{P_1P_2}$ defined by Eq.(11). The classical limit of the magnetic field effect can be taken in this case unambiguously since the FLW phase factor leads to the Schroedinger equation (14), here we have to set the  scalar potential to be zero. One obtains the classical Lorentz force and the Newton-Lorentz equation of motion from (14) as shown in Section 24 in Schiff \cite{24}. The classical correspondence is valid only if electron wavepacket is well localized and Ehrenfest theorem is applicable.

Remark two : Feynman's double slit experiment with $B=0$ is no longer a thought experiment \cite{17} as discussed in Section IIA. In the literature, double slit experiment is considered  a typical prototype quantum mechanical phenomenon. Surprisingly Feynman does not address the important question whether the quantum electron interference would survive in the presence of the local interaction with the magnetic field. He merely states that $B$ would change the positions of 'the intensity maxima and minima' in the interference pattern according to (15).

Remark three : The present paper is focused on the Aharonov-Bohm approach, however it is worth mentioning briefly the main ideas of Eherenberg and Siday \cite{1} on electron optics. Authors show that the description in terms of rays and refractive index for electrostatic focusing is quite transparent. Note that electron biprism used in Tonomura et al experiment \cite{17} is based on the electrostatic potential. Rays and wave surfaces in the case of vector potential become complicated: the refractive index is anisotropic and ${\bf A}$ being nonunique the inclination of the direction of the rays with the normal to the wave surfaces is not simple. It would be interesting to explore the geometrical optics limit in this picture to relate it with the classical correspondence, however this approach is not pursued here.

{\bf Part-B}

The ideal AB double slit electron interference setup proposed in \cite{2} is depicted in Fig.1 (Feynman's Fig.15-7). The AB phase shift (3) is nicely elucidated by Feynman. A shift in the phase difference caused by the circulation of ${\bf A}$ outside the solenoid shifts the interference pattern upwards as shown in Fig.1. Using $\delta (B=0)$ from Eq.(7) and $\delta$ given by Eq.(15) remembering that now the flux is calculated from (3) for the vector potential in the field-free region the shift in the whole interference pattern is determined denoted by $x_0$, Eq.(15-36) in \cite{14}.

Remark four : Since electron beams travel in the force free region outside the solenoid their momentum is not changed. Boyer \cite{25} emphasizes that there is no experimental or theoretical evidence for the deflection of the average momentum of the electrons. A nonclassical modular momentum exchange \cite{15} could be possible but lacks experimental support so far.

{\bf Part-C}

Feynman continues his analysis to 'show the connection between the quantum-mechanical formula and the classical formula'. A modification is suggested by him, shown here in Fig.2, that corresponds to Fig.15-8 in \cite{14} such that a constant weak magnetic field extends over a long narrow strip of width w in the region behind the slits. Ignoring for the moment the fundamental conceptual distinction between classical and quantum descriptions, let us reproduce the steps that go into the Feynman's analysis. In the first step the phase difference for the two trajectories is obtained for the flux $Bwd$ given by Feynman's Eq.15.37. The upward shift in the interference pattern is calculated from $\delta -\delta (B=0)$ to be
\begin{equation}
\Delta x =\frac{L\lambda}{hc} eBw
\end{equation}
Feynman interprets this shift 'equivalent to deflecting all trajectories' by a small angle $\alpha$ 
\begin{equation}
\alpha = \frac{\Delta x}{L}=\frac{\lambda}{hc} eBw
\end{equation}

In the second step treating electrons as Newtonian particles the Lorentz force due to the magnetic field adjacent to the slits that was neglected in the first step is considered. In the impulse approximation that this force lasts for a time $w/v$ the change in the transverse momentum is obtained to be
\begin{equation}
\Delta p_x =\frac{ewB}{c}
\end{equation}
and the corresponding angular deflection is given by
\begin{equation}
\alpha^\prime = \frac{\Delta p_x}{p}
\end{equation}
Using de Broglie relation $\lambda =h/p$ and substitution of (19) in (20) immediately give
\begin{equation}
\alpha =\alpha^\prime
\end{equation}
The exact equivalence (21) seems impressive enough for Feynman to conclude classical and quantum equivalence. 

{\bf Critique}

Part-C of Feynman's analysis has been controversial since beginning as noted by Boyer \cite{25}. A thorough critique and resolution of the controversy would be of great value in the light of current research activities \cite{3,6,7,11,12,13}. Preceding remarks on the first two parts of Feynman's account are important to delineate the crucial misleading issues.

A careful study of the complete Feynman treatment shows that there exist two fundamental weaknesses related with the quantum interference and ambiguity in defining the quantum phases. Let us first discuss the main source of confusion in the literature due to the term AB phase. Feynman does not use this term, however the physical situation depicted in Fig.1 can unambiguously be identified with AB phase and AB effect as pointed out in Part-B. Assuming logical development of his ideas proposing modification in Fig.2 and the calculation of quantum phase, his Eq.15.37, it would appear that the phase for the enclosed flux $Bwd$ is calculated using AB phase (3). Mathematical statement of the Stokes theorem (4) equating loop integral of the vector potential along a curve C to the surface integral of the magnetic field does not distinguish whether the integration path C of ${\bf A}$ lies in the magnetic field region or in the field-free region. Physically, however the two cases are entirely different: in the former case electrons experience force due to the magnetic field which is absent in the later case. We have explained in Section IIB that the term FLW phase physically distinct from AB phase should be used for the former case since the momentum exchange is a real effect in this case. Unfortunately Eq.15.37 derived in the first step by Feynman, see Part-C, remains the same irrespective of the fact whether one uses FLW phase or AB phase. This ambiguity allows an alternative interpretation of Feynman's modification such that the electron beams in two slits do experience the force. Would there still exist the interference pattern when momentum changes occur? Feynman does not address this question, see Remark two.

Double slit quantum interference is the basic framework adopted by Feynman, however the quantum mechanical nature of the interference phenomenon is only briefly discussed by him for his Fig.15-5. In the ideal AB setup of Fig.1 the absence of local field interaction and the momentum transfer to the electrons justifies his account that considers the shift of the whole interference pattern, see Remark four. In general, the presence of the fields in the slits would give rise to the comlexities in the quantum interference implied by many studies that consider the role of momentum exchange \cite{18,19,21,23} discussed in Section IIA. Feynman's formal derivation of quantum phase difference in such cases without first establishing the existence of the interference fringes could become unphysical in certain cases.

To recognize the fallacy involved more clearly we split the Feynman's modified setup (Fig.2) into two components: Fig.3 represents the analysis given in the first step, and Fig.4 that in the second step in Feynman's approach. Following the first step expression (17) is unambiguously obtained as the AB phase shift subject to the usual assumption of perfect shielding. Derivation of the angular deflection (18) has only symbolic value if it is interpreted to denote the angular shift of the interference pattern. However one cannot use the classical notion of electron trajectories. Recall the discussion in the preceding section that Tonomura et al experiment specifically demands no which-path information otherwise the interference pattern gets destroyed. The recorded positions on the screen do represent localized particle nature of the electrons but these are not determined by the classical trajectories. Thus the angular deflection $\alpha$ has no physical significance in this context.

Interestingly one can seek classical correspondence for the physical case shown in Fig.4, as well as one can treat this case in a purely classical manner. Feynman's derivation of angular deflection using Lorentz force would be correct but then it describes classical motion of electrons in a magnetic field. Mere use of de Broglie relation to transform the angular deflection $\alpha^\prime$ to the physical case of quantum interference phase shift is incorrect. Quantum mechanically one can treat this problem as that of electrons having local interaction with the magnetic field: one can use FLW phase factor for its description.

Naturally the question arises if the interference fringes could still be observed in the double slit experiment of Fig.4 when this classical limit is taken. Feynman does not address this important question. In a simple picture one would expect that in the physical situation that represents classical electron motion the quantum features would be lost. We point out a simple but profound point in this connection. In the double slit experiment the Copenhagen interpretation \cite{23} involves complementarity principle or Heisenberg uncertainty relation to explain the quantum mechanical nature of the interference phenomenon. If there is a momentum transfer during the passage through a slit then it has to be less than $h/d$ for the fringes to exist. Now the transverse momentum shift in Eq.(19) must satisfy this inequality
\begin{equation}
\Delta p_x <\frac{h}{d} ~\Rightarrow ~ wBd<\frac{hc}{e}
\end{equation}
Since the smallest Aharonov-Bohm flux unit is $hc/e$, the constraint (22) puts severe limitation on the physical validity of Feynman's second step as the fringes would be wiped out in the presence of the magnetic field.

The main results of our discussion could be summarized as follows. 1) Feynman's analysis does not establish the classical analog of the AB effect, and the classical reality of the vector potential in the modified double slit experiment. 2) The physical situation represented  exclusively by Fig.4 is interesting. Calculation of $\alpha^\prime$ will give a value given by (18), however $\alpha$ in this case is zero as there is no magnetic flux enclosed by two trajectories. Thus the claimed equivalence (21) is physically meaningless. And, 3) The existence of the electron interference itself becomes doubtful in the presence of the magnetic field in the slits, more so when the classical limit is taken.

\section{\bf Classical perspectives}

The classical limit of a quantum system is said to be the limit $\hbar \rightarrow 0$. In some cases it is known that large quantum numbers have a classical correspondence; the simple harmonic oscillator is one of the well known examples \cite{24}. In the AB effect there are intricacies due to two typical quantum characteristics: the quantum mechanical significance of the vector potential, and quantum interference. According to \cite{2} the vector potential cannot be dispensed with in quantum theory, and it has observable significance in the AB phase. Classical perspective could be put forward that the vector potential may have physical significance representing the interaction field momentum $\frac{e}{c} {\bf A}$ in view of its appearance in the canonical momentum \cite{26}. Consider a solenoid and an electron system; even for a force-free situation electron may impart rotation to the solenoid. However the classical significance of the vector potential cannot be used to offer a classical explanation of the AB effect; Trammel is careful regarding this crucial point \cite{26}. The reason is that the observable significance of the AB effect depends on the quantum phase (9) or the AB phase factor (10), on the other hand classically one has the following quantity different than the phase 
\begin{equation}
\frac{e}{c} \oint A_\mu dx^\mu
\end{equation}
The classical expression of the energy interference term derived in \cite{27} and the time-lag effect \cite{28} suffer from the defect or inconsistency that the classical quantities are interpreted as quantum phases simply by putting $\hbar$. Note that the deterministic trajectory does not make sense in quantum theory. Boyer does discuss complementarity principle in the double slit experiment, however the technical limitations of the then reported experimental tests of the AB effect seem to have led him to argue that the complementarity is not relevant in the AB phase. In the light of Tonomura et al experiment \cite{10} and modern developments on the double slit quantum interference \cite{17,18,19} briefly reviewed in Section II the role of wave-particle duality/complementarity in the AB phase shift cannot be ruled out unless alternative interpretation other than the Copenhagen interpretation, e. g. the statistical interpretation is adopted.

Local field interaction aims at the complete elimination of the vector potential the way one could do in classical theory. Unfortunately quantum formalism purely in terms of the electromagnetic fields has not been successfully developed so far. There are nevertheless interesting ideas to understand AB effect as a local effect. Feynman's schematic arrangement shown in Fig.(4) is relevant in connection with Vaidman's thought experiment \cite{3} and the example suggested in the prelude in the comment \cite{5}. Vaidman attributes the local field due to the change in the magnetic flux caused by the encircling electron, while Aharonov et al suggest assuming a constant magnetic field. In either case the role of the mechanical aspect of the solenoid or that of Vaidman's construction is not important in the AB phenomenon. In contrast, Kang \cite{4} makes an important point that the relativistically moving electron in the real experiment \cite{10} makes the ideal shielding questionable; however for a convincing argument there has to be a detailed empirical comparison with the experimental data of \cite{10}. It is also somewhat artifical to treat fluxon as a particle with mass in Kang's formalism. 

The detailed discussion in the preceding section shows that the main problem with Feynman's analysis is that the applicability of trajectory description and local field interaction amounts to the loss of the quantum phase informaion. The debate \cite{3,5,6,7} does not take notice of this most significant aspect. Recent Becker-Batelaan experiment \cite{11} assumes great significance in the context of our arguments: the experiment demonstrates the absence of back-action and approximate dispersionless forces in the AB-like setup with macroscopic toroid, and the quantum interference signifying AB effect cannot be observed in this setup. We suggest that the import of Becker-Batelaan experiment is that in the force-free region local interactions have no classical/semiclassical effect.

The classical correspondence to the AB effect could be approached from a different point of view. Recall that in the old quantum theory Bohr-Sommerfeld quantization played an important role. The most important distinguishing feature of this procedure is that the physical picture of the classical trajectories and orbits is still valid. Applying Bohr-Sommerfeld quantum rule to the canonical momentum we are led to the Aharonov-Bohm flux quantization in the expression (4)
\begin{equation}
\oint {\bf A}.d{\bf x} =n \Phi_0
\end{equation}
For a large value of the quantum number (integer n) one would expect the classical limit corresponding to (23). The flux unit is $\Phi_0=\frac{hc}{e}$. A possible observable consequence could be the deflection of the electron trajectories in the force-free region which however is indicated to be absent in the experiment\cite{11}. We suggest the experiment with intense magnetic fields i. e. large magnetic flux, and also possibly the consideration of the transverse forces noted in \cite{11}.

\section{\bf Physical mechanism} 

Aharonov and his collaborators emphasizing nonlocality as a quantum feature with no classical analog have been trying to develop a modular variable theory since 1969 \cite{15}. In the recent paper the role of modular velocity is analyzed in the so called instantaneous aspect of the AB effect \cite{7}. A continuous physical quantity modulo its basic unit is defined to be a modular variable. Modular momentum is defined as
\begin{equation}
p_x(mod ~ p_0) = p_x -N p_0
\end{equation}
where $p_0 =h/L$, and L is a constant spatial length. Transverse modular velocity is
\begin{equation}
v_x^{mod} = v_x mod~(\frac{h}{mL})
\end{equation}
here m is the electron mass. This particular variable (26) is taken to explain the AB effect associated with the vector potential
\begin{equation}
A_y =\delta(y)~H(x) \Phi
\end{equation}
Here $H(x)$ is the Heaviside step function. From a simple analysis of this example the argument is put forward that the relative AB phase shift occurs at the instant the two wavepackets cross the x-axis on which the vector potential (27) is nonzero.

The above point with great clarity had been earlier made in \cite{16} where the question was raised whether the AB phase evolved continuously. We identify two important issues from this line of thinking. When does the AB effect occur? What is the physical mechanism responsible for the AB effect? We suggest that geometry and topology provide answer to the first question. To address the second question we propose that the angular momentum holonomy conjecture \cite{20} is applicable to the AB effect.

Topological aspect was already recognized in the original paper \cite{2}, however the rich and complex topological structure involved continues to offer new avenues. Topology is a study in the global and the continuum though discrete invariants turn out to be more useful in physics. A point charge in electrostatics is the simplest example: the flux through any closed surface of arbitrary shape and area surrounding the charge is constant; it represents the topology of a punctured 3-space. In the AB effect a circle in space encircling the magnetic flux relates with the U(1) phase of the quantum state space of the electron. In both examples, the local or the geometrical aspect too is of significance when the path dependent quantity is considered. A nontrivial geometry, for example, the surface of a 2-sphere gives rise to the change in the direction of a vector parallel transported from one point to another, and termed holonomy for a closed circuit. Geometric phases in optics are the manifestations of this holonomy \cite{20}. Altering the 2-sphere scenario with a magnetic monopole at its centre Wu-Yang analysis \cite{9} shows that the path dependent phase factor becomes undefined when the path crosses the singularity. In a construction having singularity-free potentials defined in two regions the Dirac quantization condition emerges as a topological property. Analogous to the monopole problem we associate geometrical aspect contained in the FLW phase factor for a path on the circle such that the polar angle $\theta ~ \in ~[0,2\pi)$ where the phase evolution is continuous. In the presence of the magnetic flux at the origin the topology is that of a 2-space with the origin removed. Mathematically following \cite{2} single valued wavefunctions on the two paths encircling the singularity could be used to calculate the phase difference or one could treat the problem with a multivalued wavefunction with well defined observables as shown by Martin \cite{29}. Thus the topological aspect is connected to the transition point in the path winding the circle.

To see the topological aspect more clearly let us note that the FLW phase factor is calculated by dividing the path into segments and integrating along them, however for the closed path the gauge invariance brings the gauge function $\chi$ into the picture. If the integration loop is inside the magnetic field region the boundary condition of the gauge field ${\bf A}$ is simple and continuous resulting into geometric phase with trivial topology. In the case of AB phase the crossing of the singularity gives rise to a nontrivial topology of the AB effect or we say that a lift of the loop in the covering space where the real line is the covering space of the circle. The AB phase thus evolves continuously until the discontinuity in the vector potential is crossed. Martin shows the importance of the irreducible representations of the Euclidean group E(2) and its covering groups parameterized by two real numbers for the AB effect; a lucid and comprehensive account is given by Kastrup \cite{30}.

Recognizing the necessity of a curved geometry (i. e. circular arc) and a topological defect causing a discontinuity as two elements in the AB effect we argue that the field angular momentum exchange similar to that proposed for the geometric phases in optics \cite{20}, shown to have experimental support \cite{31}, also provides a physical mechanism for this effect. Though the idea of modular momentum exchange \cite{16} is interesting \cite{13, 32} the role of angular momentum appears more natural. In fact, Aharonov and Cohen discuss rotating frame of reference and geometric effect in a recent paper \cite{33}. In this connection first we recall an elegant example \cite{34}. The Hamiltonian of a two dimensional isotropic oscillator with unit mass
\begin{equation}
H_0 =\frac{1}{2} (\dot{x}^2 +\dot{y}^2) +\frac{\omega^2}{2} (x^2 +y^2)
\end{equation}
where overdot denots the time derivative, is invariant under the rotation of the axes by an angle $\alpha$
\begin{equation}
x \rightarrow x ~ cos \alpha +y~ sin \alpha
\end{equation}
\begin{equation}
y \rightarrow - x ~sin \alpha +y ~cos \alpha
\end{equation}
The constant of motion for this symmetry is obviously the angular momentum. If $\alpha$ is made time dependent then $H_0$ is not invariant under the rotation, however a term $H_I$ added to $H_0$ makes the total Hamiltonian invariant provided the frequency changes as
\begin{equation}
\omega ~\rightarrow ~ \omega +\dot{\alpha}
\end{equation}
Here $H_I$ is given by
\begin{equation}
H_I = -\omega(\dot{x} y -\dot{y} x)
\end{equation}
The change in the angle variable after a period of time T from (31) gives
\begin{equation}
\Delta \theta =\int_0^T \omega (t) ~dt +\int^T_0 \dot{\alpha} ~dt
\end{equation}
The second term in (33) was interpreted to have topological origin in \cite{34}. This simple example illustrates the role of gauge invariance (31) and angular momentum exchange via (32).

Martin's mathematical model \cite{29} would make the argument cogent. Rotation (29)-(30) and translation for the group E(2) and its covering groups are defined by two real numbers $\rho >0; \beta \in [0,1)$. For E(2) itself $\beta =0$, and for a rational number $\beta =p/q,~$ p and q being integers with no common divisors it is a q-fold covering group of E(2). The self-adjoint generators of rotations and translations satisfy the Lie algebra \cite{30}. Of particular interest for the present discussion is the rotation generator
\begin{equation}
L_\beta = -i\frac{\partial}{\partial \theta} +\beta
\end{equation}
In quantum mechanics for the state space we require a Hilbert space on the circle. A Hilbert space with square-integrable functions and a well defined scalar product can be constructed using the orthonormal basis vectors
\begin{equation}
e_n(\theta) = e^{i n\theta} ~ ~ n \in Z
\end{equation}
Multiplying (34) by $\hbar$ and interpreting it as angular momentum operator it is easily verified that
\begin{equation}
\hbar L_\beta e_n(\theta) = \hbar (n+\beta) e_n ~ ~ n \in Z
\end{equation}
The functions (35) are the eigenvectors of the angular momentum operator with noninteger eigenvalues $\hbar(n+\beta)$. The nontrivial topology has the origin in the homotopy group of the nonsimply connected circle
\begin{equation}
\Pi_m(S^1) = Z,~ ~ m=1 ; 0, ~ ~m>1
\end{equation}

The number $\beta$ determines the different irreducible representations in the same Hilbert space defined by the basis vectors (35). Using the unitary transformations on (35) separate Hilbert spaces for each $\beta$ could be constructed, and the operator (34) becomes independent of $\beta$
\begin{equation}
\hbar L_\beta = - i \hbar \frac{\partial}{\partial \theta}
\end{equation}
The basis vectors now have a constant multiplicative phase factor $e^{i 2\pi \beta}$. Note that the eigenvalues do not change under the unitary transformation.

It is remarkable that the eigenspectrum (36) of the angular momentum operator resembles the definition of a modular variable, e. g. the modular momentum defined by Eq.(25). A new interpretation of the AB effect emerges in terms of the modular angular momentum exchange. Consider the flux line following \cite{2} in cylindrical coordinates and Coulomb gauge the vector potential is
\begin{equation}
A_r = A_z =0  ~ ~A_\theta = \frac{\Phi}{2\pi r}
\end{equation}
The dynamics of an electron is described by 
\begin{equation}
H \Psi = i \hbar \frac{\partial \Psi}{\partial t}
\end{equation}
\begin{equation}
H= -\frac{\hbar^2}{2 m} [ \frac{\partial^2}{\partial r^2} +\frac{1}{r} \frac{\partial}{\partial r} -\frac{1}{r^2} (-i\frac{\partial}{\partial \theta} + \gamma)^2]
\end{equation}
where $\gamma =-\frac{e\Phi}{c h}$. The Casimir invariant is just $\frac{2m}{\hbar^2} H$, and the angular momentum operator becomes
\begin{equation}
L_{AB} =-i \hbar \frac{\partial}{\partial \theta} +\hbar \gamma
\end{equation}
A simple calculation comparing (42) with (34) and noting the phase factor $e^{i 2\pi \gamma}$ the expression for the phase shift $2\pi \gamma$ is exactly the AB phase shift (3).

To gain further insight into the role of angular momentum in the AB effect we recall the arguments in \cite{20}. The crucial difference between the kinetic and canonical momenta in the classical point dynamics manifests in a general form in the field theory \cite{35}. For free electromagnetic field the canonical angular momentum density tensor $M^{\alpha \mu \nu}$ derived as a Noether current for the invariance of the action under infinitesimal proper homogeneous Lorentz transformation is not gauge invariant. A gauge invariant tensor $J^{\alpha \mu \nu}$ can be constructed that differs from the canonical one by a pure divergence term
\begin{equation}
M^{\alpha \mu\nu} = J^{\alpha\mu\nu} -\partial_\lambda [F^{\lambda\alpha} (A^\mu x^\nu -A^\nu x^\mu)]
\end{equation}
In the standard field theory \cite{35} it is assumed that the divergence term vanishes in the limit of rapidly falling fields at infinity. In a nontrivial geometry this assumption may no longer remain valid and the surface terms could result into what we termed as angular momentum holonomy \cite{20}.

The 4-vector potential term in the expression (43) is interesting
\begin{equation}
A^\mu x^\nu -A^\nu x^\mu
\end{equation}
Its 3-space analogue is just the term ${\bf r} \times {\bf A}$. This term also arises from the expression for the kinetic momentum ${\bf r}\times ({\bf p}-\frac{e}{c} {\bf A})$ considered by Trammel \cite{26}. The angular momentum operator (42) is also related with the gauge covariant derivative. Therefore the idea of modular angular momentum exchange seems quite logical.

Instead of a flux line having singularity, let us consider a solenoid with finite radius R and a constant magnetic field $B_0$ along z-axis. The physical situation for this case is represented by Fig.1. Electrons move in the field-free region, i. e. outside the solenoid where the magnetic field is zero. The vector potential inside and outside regions is
\begin{equation}
A^{in}_r =A^{in}_z =0 ~~ A^{in}_\theta =\frac{B_0 r}{2}
\end{equation}
\begin{equation}
A^{out}_r =A^{out}_z =0 ~~ A^{out}_\theta =\frac{B_0 R^2}{2 r}
\end{equation}
It can be seen that the vector potential in the field-free region ${\bf A}^{out}$ is a pure gauge potential
\begin{equation}
{\bf A}^{out} = {\bf \nabla} (\frac{B_0 R^2 \theta}{2})
\end{equation}
The quantity $\frac{e}{c} {\bf r}\times{\bf A}$ for the two regions is calculated to be
\begin{equation}
L^{in}_z = \frac{eB_0 r^2}{2c} = \frac{e\Phi}{2\pi c} \frac{r^2}{R^2}
\end{equation}
\begin{equation}
L^{out}_z =\frac{eB_0 R^2}{2c} =\frac{e\Phi}{2\pi c}
\end{equation}
where $\pi R^2 B_0 =\Phi$. Interestingly using the flux quantum unit $hc/e$ the expression (49) can be transformed to the modular form
\begin{equation}
L^{out}_z (mod ~ \hbar) =L^{out}_z -n \hbar
\end{equation}

To picture the physical process of modular angular momentum exchange first it has to be realized that electrons in the field-free region do not undergo velocity change. However the motion on a curved trajectory with constant velocity brings the role of angular momentum and apparent forces. Now the pure gauge potential, in general, could be viewed as a fictitious magnetic field \cite{36} using a vector identity. Let ${\bf A} =-{\bf \nabla} \eta$ then we have the following equality
\begin{equation}
{\bf r} \times {\bf A} = {\bf \nabla} \times {\bf r} \eta ={\bf B}_{fic}
\end{equation}
Expression (51) provides alternative argument to explain the effect of the confined magnetic flux on the electron motion in the field-free region since the vector potential (47) is a pure gauge potential.

\section{\bf Conclusion}

Recent experiment \cite{11} and the continued efforts seeking the classical limit of the AB effect by many authors motivated us to present a comprehensive discussion on the classical perspective. For this purpose, a critique on the Feynman approach is offered and it is concluded that the AB effect has no classical analog consistent with the original claim \cite{2}.

The idea of modular variables \cite{16} seems interesting, however based on topological arguments we suggest modular angular momentum exchange as a physical mechanism for the AB effect. The present work provides support to the speculation that potentials are fundamental in the microscopic domain. A typical gauge chosen in \cite{7,16} described by the vector potential (27) to explain the instantaneous aspect of the AB effect is not discussed here for the lack of the definitive result.

The issue of the physical significance of canonical versus kinetic quantities is a delicate one. We refer to the recent literature \cite{37} and the references cited therein. The present work may prove to be useful in this context.

{\bf Captions~ to ~Figures}

Fig.1 Schematics of the Aharonov-Bohm double slit setup.

Fig.2 Feynman's modification in the AB setup.

Fig.3 Splitted part of the Feynman's schematic analogous to the AB setup.

Fig.4 Second part of the Feynman's modified setup depicting the local interaction of electrons.

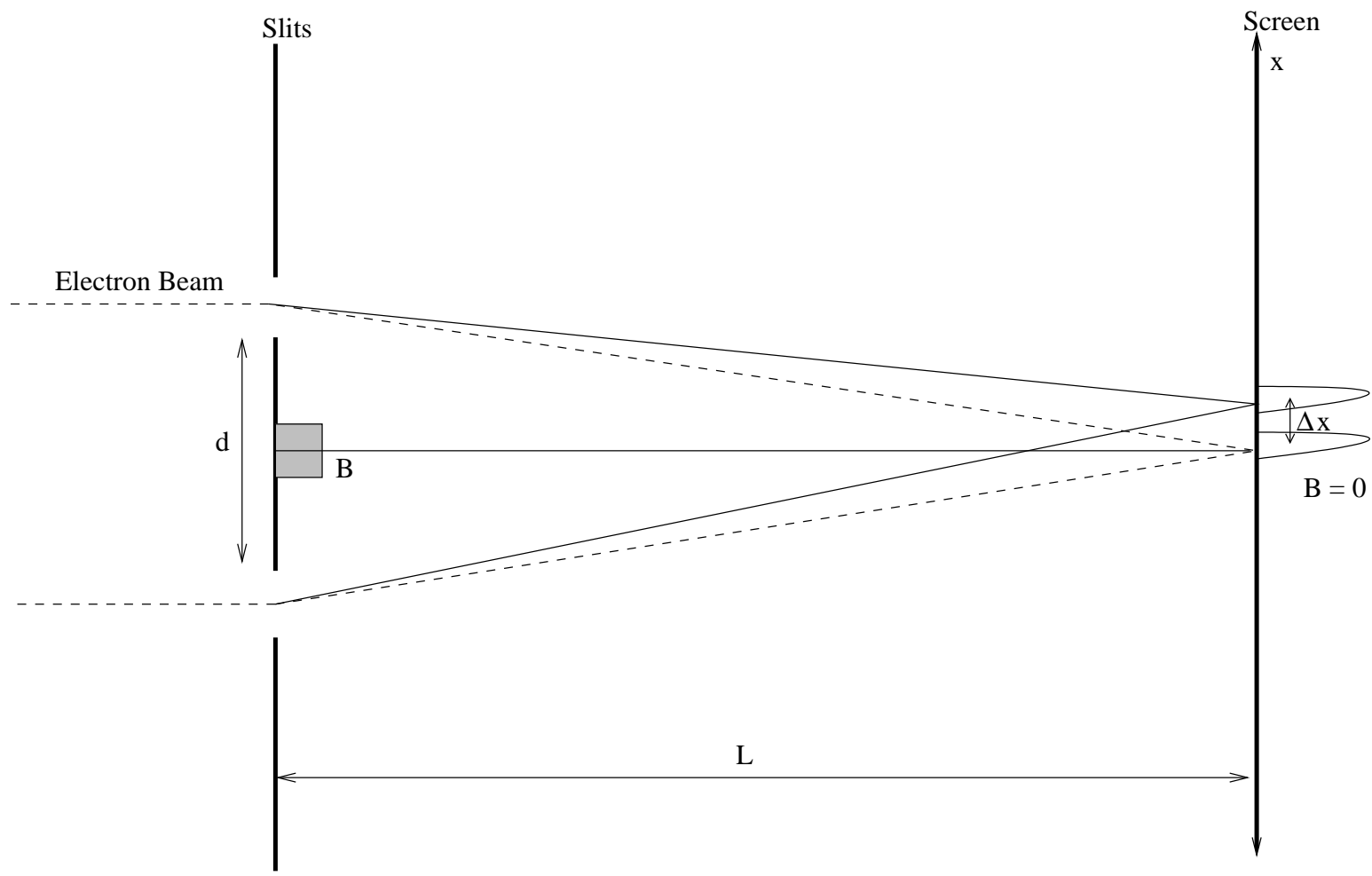
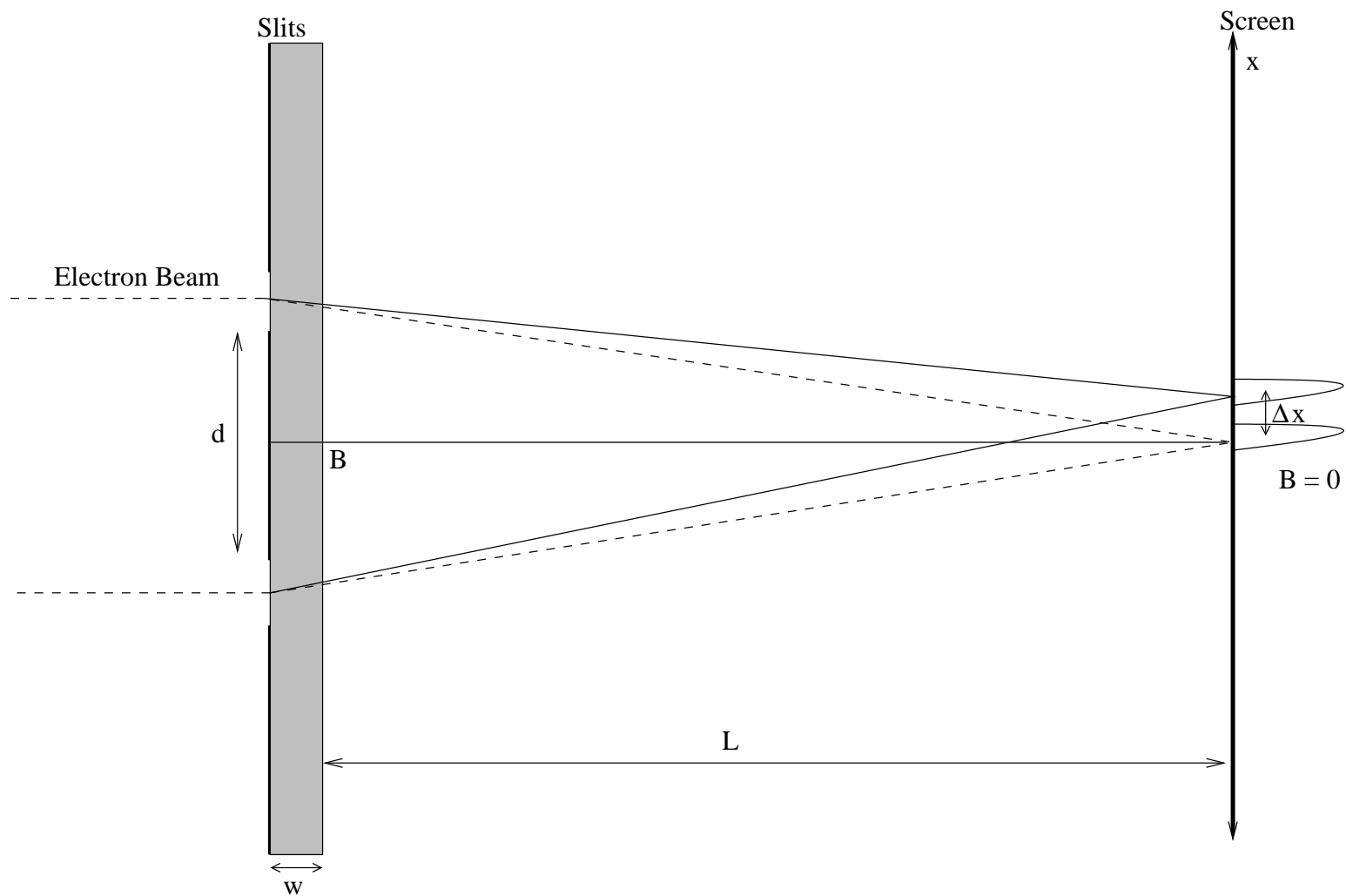
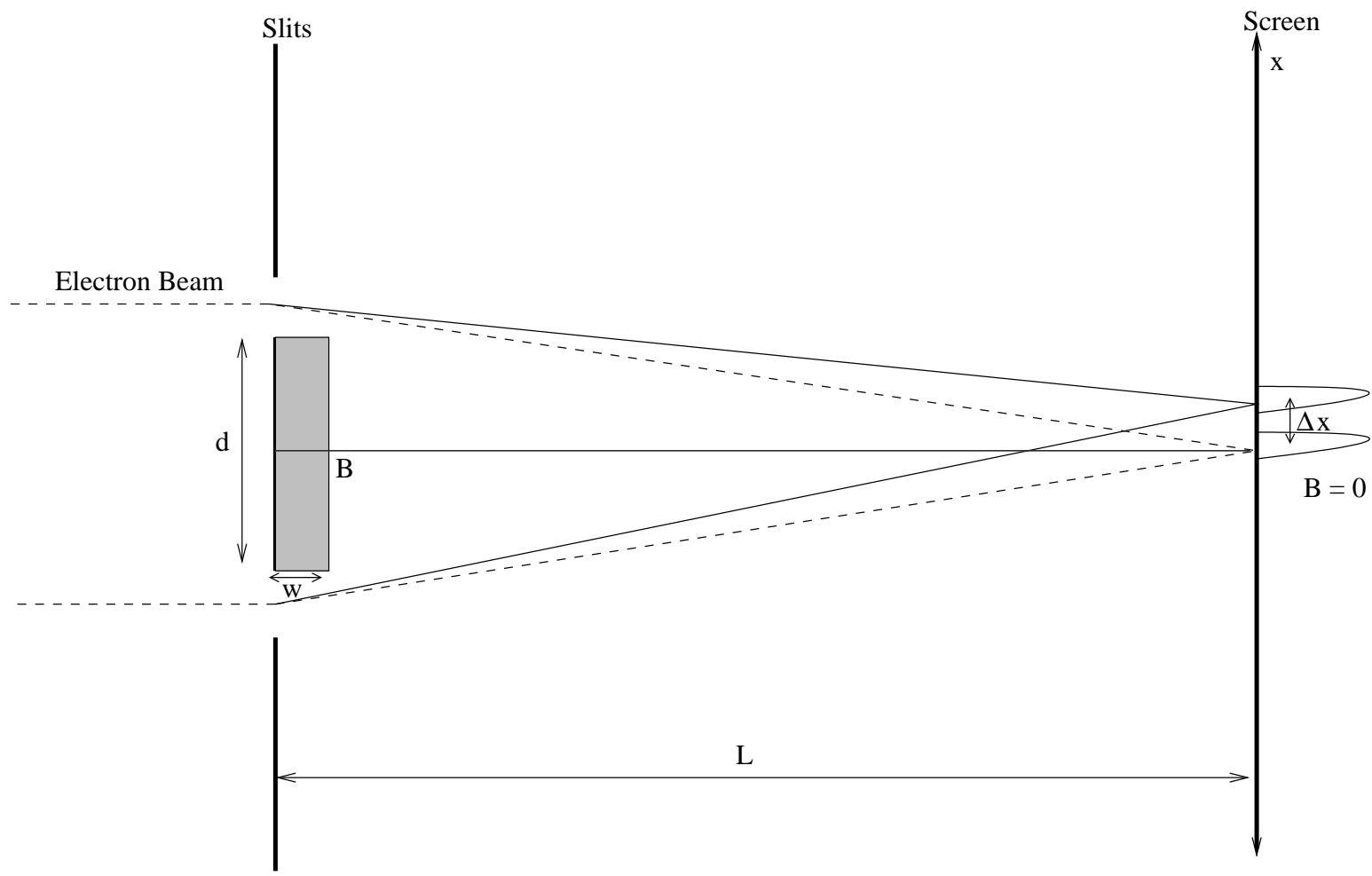
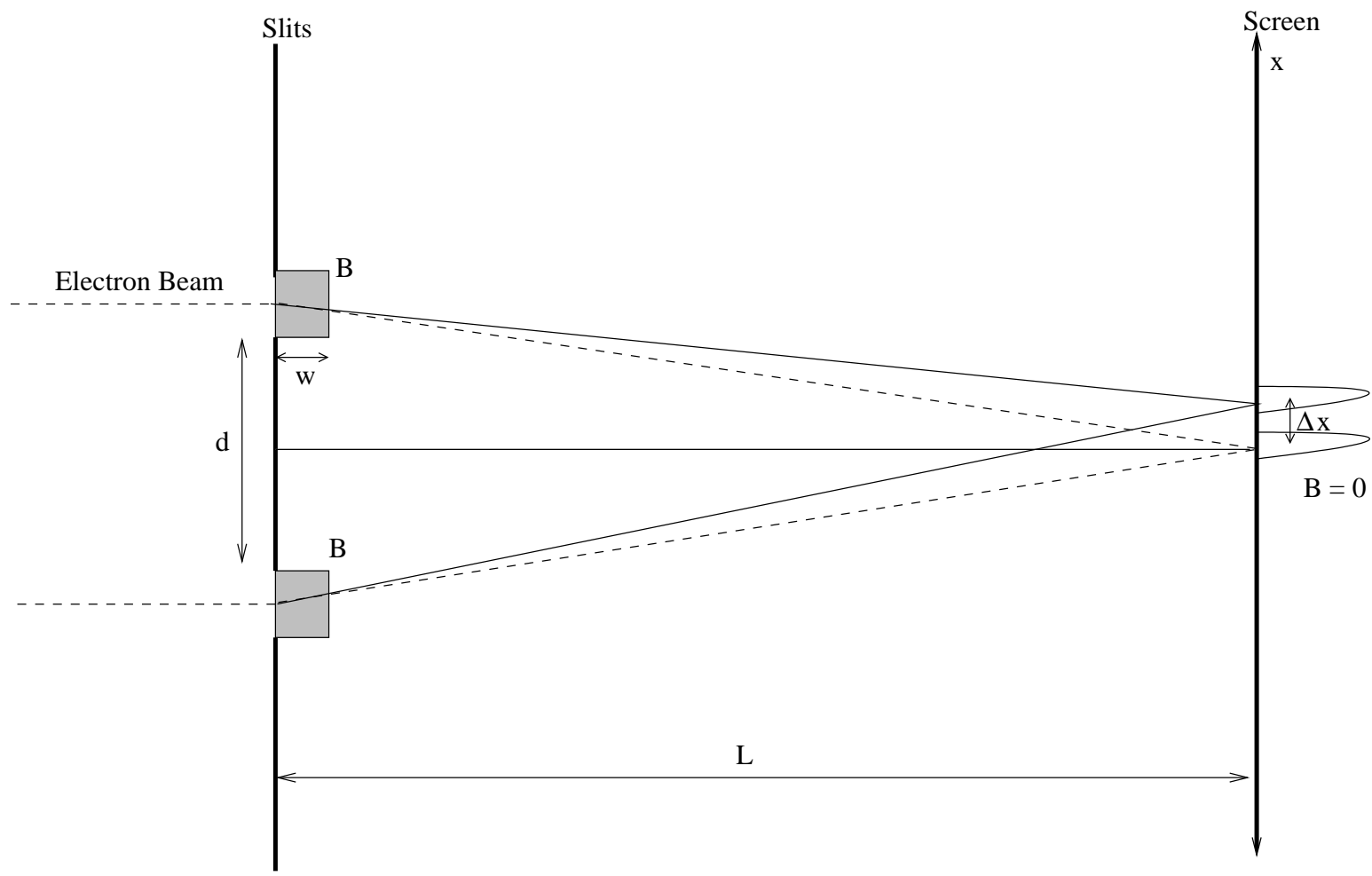
\end{document}